# Size and temperature dependent magnetization of iron nanoclusters

G. Dos Santos,[1, 2] R. Aparicio,[1, 2] D. Linares,[3] E.N. Miranda,[4] J. Tranchida,[5] G.M. Pastor,[6] and E.M. Bringa[1, 2, 7]

[1]*CONICET, Mendoza, 5500, Argentina*
[2]*Facultad de Ingeniería, Universidad de Mendoza, Mendoza, 5500 Argentina*[*]
[3]*Departamento de Física, Instituto de Física Aplicada, Universidad Nacional de San Luis-CONICET, Ejército de Los Andes 950, D5700HHW, San Luis, Argentina*
[4]*IANIGLA-CONICET, CCT Mendoza, 5500-Mendoza, Argentina*
[5]*Computational Multiscale, Center for Computing Research, Sandia National Laboratories*
[6]*Institute of Theoretical Physics, University of Kassel, 34134 Kassel, Germany*
[7]*Centro de Nanotecnología Aplicada, Facultad de Ciencias, Universidad Mayor, Santiago, Chile 8580745*

(Dated: August 28, 2020)

The magnetic behavior of bcc iron nanoclusters, with diameters between 2 and 8 nm, is investigated by means of spin dynamics (SD) simulations coupled to molecular dynamics (MD-SD), using a distance-dependent exchange interaction. Finite-size effects in the total magnetization as well as the influence of the free surface and the surface/core proportion of the nanoclusters are analyzed in detail for a wide temperature range, going beyond the cluster and bulk Curie temperatures. Comparison is made with experimental data and with theoretical models based on the mean-field Ising model adapted to small clusters, and taking into account the influence of low coordinated spins at free surfaces. Our results for the temperature dependence of the average magnetization per atom $M(T)$, including the thermalization of the transnational lattice degrees of freedom, are in very good agreement with available experimental measurements on small Fe nanoclusters. In contrast, significant discrepancies with experiment are observed if the translational degrees of freedom are artificially frozen. The finite-size effects on $M(T)$ are found to be particularly important near the cluster Curie temperature. Simulated magnetization above the Curie temperature scales with cluster size as predicted by models assuming short-range magnetic ordering (SRMO). Analytical approximations to the magnetization as a function of temperature and size are proposed.

## I. INTRODUCTION

At the nanoscale, finite-size effects can strongly influence the magnetic properties of materials [1]. Fe layers deposited on W substrates (typically on the order of few hundreds of Fe atoms) are a prototypical example of those effects [2]. Numerical and experimental studies have extensively demonstrated the impact of the size, dimension and number of Fe monolayers on their magnetic properties, including ordering temperature (Curie or Néel), magnetic susceptibility or magnon dispersion relations [3–5]. In the field of magnetic nanoclusters, large departures from the corresponding bulk magnetic properties have also been observed. For example, hysteresis loop, coercive field, ordering temperature or spontaneous magnetization, have been shown to drastically depend on the size of iron oxide nanoparticles [6–8].

Understanding magnetism at the nanoscale is important since the computed or measured magnetic properties can be used to parametrize micromagnetic models, which are extremely valuable in order to simulate technological applications [9]. This is the case, for example even in the most basic Stoner-Wohlfarth (SW) model [10], which represents the coercivity and switching field of a magnetic single-domain nanoparticle. In its simplest form, the SW model depends on the total magnetization and anisotropy energy of the particle. It is therefore of fundamental importance to develop accurate numerical tools evaluating how the cluster magnetization is affected by temperature and particle size. Electronic *first-principles* calculations are certainly extremely valuable to obtain better insight at localized effects and to derive magnetic interaction parameters [11]. However, the involved computational costs and their poor scalability makes them unpractical for simulating nanoparticles in the size range of technological interest, as they are typically limited to systems up to few hundreds of atoms [12, 13]. Consequently, developing reliable novel approaches is crucial for the progress in this field.

Leveraging an adiabatic atomistic spin approximation [14], Atomistic Spin Dynamics (ASD) is a widely used classical spin-lattice methodology allowing to model complex nanoscale systems [15–18]. Magnetic trajectories are simulated on a potential energy surface generated by a magnetic Hamiltonian usually parametrized from *first-principles* calculations [19]. Assuming fixed atomic positions, ASD simulations do not account for magnon-phonon interactions. Former studies have displayed the importance of those interactions on the description of materials properties such as magnon lifetime, phononic thermal conductivity, magnetic switching or critical temperature [20–22]. Spin dynamics (SD) can also be used as a coarse grained approach, to compute the time evolution of blocks of the systems having many atoms with some effective magnetization. This allows the simulation of micron-sized systems that are of technological relevance [23–26]. ASD has been recently applied to Fe oxide nanoparticles (NPs), using a triangular lattice [27].

[*] gonzalodossantos@gmail.com

A study accounting for the magnon-phonon interactions in molecular dynamics (MD) simulations has been performed by Dudarev and Derlet. Using a combination of the Stoner and the Ginzburg-Landau models, they developed a "magnetic" potential for $\alpha$ iron in order to take into account some effects of magnetism, including the energetics of point defects [28, 29]. Later on, a numerical methodology coupling MD and ASD by explicitly treating atomic and spin degrees of freedom as well as their coupling through a magneto-elastic Hamiltonian, was presented by Ma *et al.* [30]. During the last few years, a large number of investigations have been carried out applying molecular dynamics coupled to spin dynamics simulations (MD-SD) in order to explain experiments of magnetic instability [31], demagnetization, impact of temperature on magneto-mechanical properties and phase transitions [21, 32]. This includes the development of a new software for the implementation of the model, SPILADY [33]. Perera *et al.* have carried out studies of magnetic Fe using MD-SD [34] including spin-orbit coupling effects [35]. Other recent studies incorporate additional exchange parameters obtained from *ab-initio* methods [36].

In this work spin dynamics coupled to classical molecular dynamics simulations [37] is used to incorporate thermal spin and mechanical effects which are difficult to include in ASD simulations. This method is applied to Fe nanoclusters to obtain magnetization versus size and temperature, subsequently comparing those results to semianalytical models.

The paper is organized as follows. In section II the simulation framework employed is presented, as well as the details of the calculations. In section III, two semi-analytical models are introduced in order to qualitatively analyze numerical simulations. The results are presented and discussed in section IV. Finally, the main conclusions are drawn in section V.

## II. METHODS

### A. Theoretical framework

In this work we perform MD-SD, where the spin degrees of freedom are coupled to the lattice degrees of freedom. For this purpose, we run our simulations under the SPIN package recently added to the software LAMMPS [37]. Under this framework one is able to introduce magnetic effects in a classical MD simulation through a generalized Hamiltonian,

$$\mathcal{H} = \sum_{i=1}^{N} \frac{|\boldsymbol{p}_i|^2}{2m_i} + \sum_{i,j,i \neq j}^{N} V(r_{ij}) + \mathcal{H}_{\text{mag}}. \quad (1)$$

The first term is the kinetic energy of the atoms and the second term is a classical interatomic potential describing the mechanical interactions between the atoms. The last term is a magnetic Hamiltonian, which in its simplest form is given by an exchange Heisenberg interaction:

$$\mathcal{H}_{mag} = -\sum_{i,j,i \neq j}^{N} J(r_{ij}) \boldsymbol{s}_i \cdot \boldsymbol{s}_j. \quad (2)$$

This Hamiltonian term represents the pair interaction between spins, where $\boldsymbol{s}_i$ is the normalized spin vector of spin $i$ and $J(r_{ij})$ is the Heisenberg magnetic coupling exchange constant, which depends on the distance $r_{ij}$ between atoms $i$ and $j$. Other terms, including anisotropy and dipolar terms, can be added to the Hamiltonian. The distance dependence of the exchange constant $J(r_{ij})$ is a key aspect of the model since it mediates the spin and lattice coupling. Furthermore, $J(r_{ij})$ is modelled by a function based on the Bethe-Slater curve [38, 39], parameterized using three coefficients that must be fitted to ab-initio calculations:

$$J(r_{ij}) = 4\alpha \left(\frac{r_{ij}}{\delta}\right)^2 \left[1 - \gamma \left(\frac{r_{ij}}{\delta}\right)^2\right] e^{-\left(\frac{r_{ij}}{\delta}\right)^2} \Theta(R_c - r_{ij}) \quad (3)$$

where $\Theta(R_c - r_{ij})$ is the Heaviside step function and $R_c$ is the cutoff distance. In the present work we have used the parametrization for bcc iron described in previous works [20, 37] from ab-initio calculations reported by Pajda *et al.* [40]. Specifically, the values of the fitting parameters are $\alpha = 25.498$ meV, $\gamma = 0.281$, $\delta = 0.1999$ nm, and exchange interaction cutoff distance $R_c = 0.34$ $nm$. The same $J(r_{ij})$ applies to all atoms in the NP. Notice that there is no general agreement about the exact value of the coupling-exchange constant for different interatomic distances for bcc Fe. In fact, the reported values of $J(r_{ij})$ from ab-initio calculations found in the literature show large discrepancies [40–42].

Temperature in other simulation schemes, such as micromagnetic simulations, is not uniquely defined and it has to be rescaled in order to compare with experiments [24]. In the simulations presented here lattice temperature has a clear definition, and magnetic temperature is defined as in Tranchida *et al.* [37]. In addition, in these spin-lattice simulations the effect of temperature in lattice expansion is more realistic (due to the spin-lattice coupling) than in some previous approaches, which consider for example an homogeneous linear thermal expansion of the lattice with fixed atoms [43].

The interaction between atoms is modeled using an embedded atom model (EAM) interatomic potential [44] which describes reasonably well several Fe properties, including the thermal expansion, phonon dispersion curves, mean-square displacements and surface relaxations. The interatomic cutoff distance for this potential was set to 0.35 nm.

The simulations are performed using classical atomic dynamics and classical spin dynamics, without considering any quantum-mechanical effects. This is in line with the adiabatic approximation. However, the behavior at cryogenic temperatures might not be well described.

Moreover, other effects resulting from electron-phonon and electron-spin coupling are not included. There are studies which include electrons within a "classical" two-temperature model (TTM) approach for electrons and atoms, leading to a 3-temperature model when spins are also included [33].

The spin dynamics is calculated using a Landau-Lifshitz-Gilbert (LLG) approach [14], integrated with a Suzuki-Trotter integrator. Details can be found in Tranchida *et al.* [37].

## B. Simulation details

Body centered cubic (bcc) iron NPs are simulated ranging their diameter and temperature from 2 to 8 nm and from 10 to 1300 K, respectively. In addition, "bulk" simulations were run and used as a reference in order to address the finite-size effects of the NPs. In these cases, we have modeled cubic simulation boxes with linear sizes of $10\,a_0$, $20\,a_0$ and $30\,a_0$ (with $a_0 = 2.8665$ Å the bcc Fe lattice parameter) containing 2000, 16000 and 54000 atoms respectively, and considering periodic boundary conditions in the three directions.

The magnetization values for the bulk systems were obtained from linear finite-size scaling analysis. As shown in Fig. 1, the magnetization values for the systems with $L = 10$, 20 and 30 are plotted as a function of $1/L$, and then the bulk magnetization is determined as the intercept of a linear fit for those points.

All systems were simulated without external magnetic field, and considering cubic magnetic anisotropy along the 3 main axis of the bcc lattice (details of how the anisotropy is included in the simulations are given in [37]). The anisotropy constants were set to $K_1 = 3.5\mu$eV/atom and $K_2 = 0.36\mu$eV/atom [45]. It has been argued that anisotropy values vary near the surface of nanoparticles due to the reduction of coordination [46, 47]. Nevertheless, as shown by Ellis *et al.* [48], for example, the overall effect in FePt nanoparticles of 4.632nm and 2.316nm is a change of about 10% in the anisotropy constant. In this work, the anysotropy constant is assumed to be the same for all nanoparticles, despite surface proportion being clearly different for the sizes studied here.

Initially, all atomic spins were oriented along the z axis ([001] direction). This initial configuration was chosen since the magnetization reaches an equilibrium value faster than when the spins start from a random configuration (see Fig. S1 in the supplementary material).

All simulations span more than 0.5 ns, using a timestep of 1.0 fs. All samples were initially thermalized to equilibrium using a Langevin thermostat applied to the lattice and another thermostat applied to the spins in order to ensure fast thermal equilibrium. In all cases we have used a damping factor of $\lambda_L = 1.0$ for the lattice thermostat and a transverse damping $\lambda_s = 0.1$ for the spin system. Before the thermostat was applied, atomic velocities were set so as to obtain an initial temperature of 300 K (or 10 K in the cases of the simulations with T<300 K). Applying the thermostat, thermal equilibrium was reached in all cases in a time between 2 ps and 10 ps with some fluctuations around the set temperature. Similarly, the resulting magnetization quickly reaches a stable value and remains stable during the simulation although its components may fluctuate, as shown in Fig. 2.

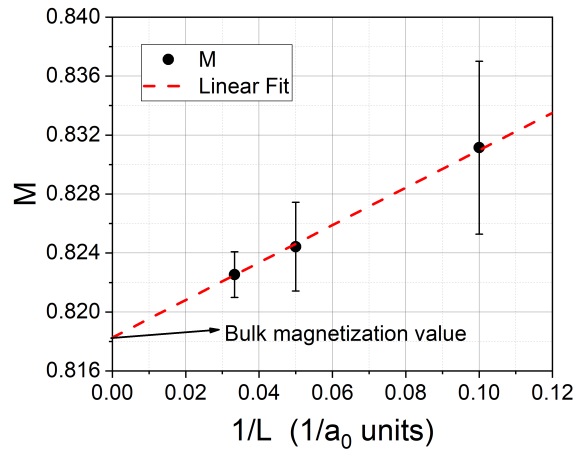

Figure 1. Finite-size scaling for bulk magnetization simulations, using periodic boundary conditions. The data correspond to simulations run at 300 K and error bars show standard deviation resulting from averaging magnetization over the final 0.5 ns.

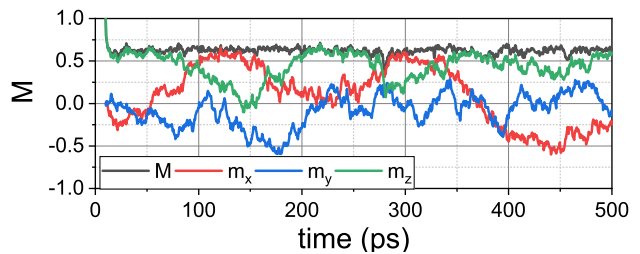

Figure 2. Time dependence of the components $m_x$, $m_y$, $m_z$, of the magnetization along the $x$, $y$ and $z$ axes and the total magnetization $M$ during a typical simulation. The results correspond to simulations of a 2nm wide sphere at 400 K.

Once the simulation results were obtained, we conducted post-processing analyses using the free software Ovito [49]. In particular, we employed the coordination analysis tool to distinguish between atoms in the outer layer of the sphere ("shell" or "surface") and the inner ones ("core") in order to obtain the magnetization curves of these two different regions. In all cases a single layer of atoms was selected within the surface group, as shown in Fig. 3. Details of the number of atoms contained in each NP and the ones belonging to the surface and the core regions are presented in Table I.

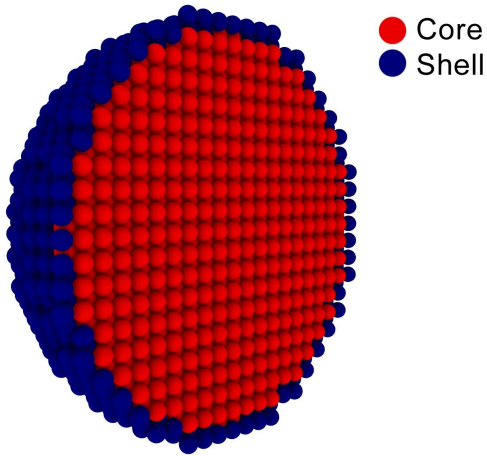

Figure 3. Snapshot of one half of a 6nm diameter Fe NP, obtained with software OVITO [49], inscribed in a cubic region with periodic boundary conditions. The shell and core regions are indicated by the colors, where each small sphere represents an atom.

Total magnetization averages were computed over the last 300000 steps of the simulation. To minimize computational cost, core and shell magnetization ($M_c$ and $M_s$ respectively) results were obtained averaging over the last 100 ps of the simulation.

| Diam. | # of atoms | Surf. atoms | Core atoms | % surface | % core |
|---|---|---|---|---|---|
| 2nm | 339 | 177 | 162 | 52.21 | 47.79 |
| 3nm | 1243 | 528 | 715 | 42.48 | 57.52 |
| 4nm | 2741 | 808 | 1933 | 29.48 | 70.52 |
| 5nm | 5601 | 1480 | 4121 | 26.42 | 73.58 |
| 6nm | 9577 | 2054 | 7523 | 21.45 | 78.55 |
| 8nm | 22659 | 3695 | 18964 | 16.31 | 83.69 |

Table I. Total number of atoms, surface atoms, core atoms, and percentage of surface and core atoms for each studied Fe nanocluster.

## III. MEAN-FIELD ISING MODELS

It is interesting to compare and contrast the results found in the numerical simulations by using two variants of the mean-field Ising model.

### A. Spin model including surface effects

Our results can be compared with a theoretical model that includes surface effects on the magnetization. This model, originally proposed by Mills [50], is known as semi-infinite Ising model with a free surface, i.e. basically a mean-field Ising model of a ferromagnet with a free surface. It is assumed that the spins are arranged in a lattice (bcc in this case) and that the spins in all lattice sites are given by $S_i = \pm 1$ where $S_i = +1$ means that spin $i$ is pointing along the positive $z$ direction and $S_i = -1$ means that spin $i$ is pointing in the opposite direction. In this model the exchange coupling constant $J$ is the same for all nearest-neighbor pairs, except for the case of pairs at the surface of the ferromagnet where it is denoted by $J_s$. A layered structure is then obtained with the Hamiltonian being given by

$$\mathcal{H} = -J \sum_{\langle i,j \rangle \notin S}^{N} \boldsymbol{\mu}_i \cdot \boldsymbol{\mu}_j - J_s \sum_{\langle i,j \rangle \in S}^{N} \boldsymbol{\mu}_i \cdot \boldsymbol{\mu}_j. \quad (4)$$

Following the mean-field approach, the magnetization for the surface $m_s$ and for each successive layer $m_1, m_2, \ldots m_n$, are given by,

$$m_s = \langle \mu_{i \in S} \rangle = \tanh(4K_S m_S + K m_1) \quad (5)$$
$$m_1 = \langle \mu_{i \in 1} \rangle = \tanh(4K m_1 + K m_S + K m_2) \quad (6)$$
$$m_n = \langle \mu_{i \in n} \rangle = \tanh(4K m_n + K m_{n-1} + K m_{n+1}) \quad (7)$$

where $K_s = \beta J_s$ and $K = \beta J$ with $\beta = 1/k_B T$ and $k_B$ is the Boltzmann constant. It is important to notice that $K$ and $K_s$ are not related to the anisotropy constants $K_1$ and $K_2$ described previously. In this work we have adopted a two-layer approximation, meaning that we have considered that the system's magnetization is unaltered after the 1st layer (the one after the surface), $m_1 = m_2 \cdots = m_n = m_{bulk}$. Therefore, $m_1$ represents the cluster core magnetization and, in addition we have set $J_s = J$. In this way, the summation on the first term of the right-hand side of the Hamiltonian (Eq. 4), runs over the 8 first neighbours (core) and the one on the second term runs over 4 first neighbours [(100)-like surface]. Related electronic models have been used to obtain the magnetization of Fe thin films [43].

### B. Spin model including finite-size effects

A previous theoretical model developed in Ref. [51] is summarized below. The model is based on the mean-field approximation for the Ising model, adapted to the statistics of few-particle systems. The mean-field approach is well known and does not need explanation. However, if a few-particle system is analyzed we should be careful with the approximations that are used when obtaining the fundamental equations in the microcanonical formalism [52]. In particular, the Stirling approximation ($\ln x! \approx x \ln x - x$) cannot be applied, and the factorial must be written in terms of the Gamma function: $\Gamma(x) = (x+1)!$. This implies working with the derivative of this function, known as the digamma function

$$\psi^{(0)}(x) = \frac{d\Gamma}{dx} \quad (8)$$

Taking this into account, a self-consistent equation for the magnetization M is found. If there are N atoms in the system and the coordination number between them is z, the magnetization is given by the solution of the equation

$$M = \frac{k_B T}{zJ}\left\{\psi^{(0)}\left[\frac{N}{2}(1+M)+1\right] - \psi^{(0)}\left[\frac{N}{2}(1-M)+1\right]\right\} \quad (9)$$

Naturally, this simple model does not allow us to make precise quantitative predictions due to the coarseness of the mean-field approach. However, one can expect to obtain an interesting qualitative comparison with the results of the simulations. Eq. (9) has to be solved numerically for the conditions in each simulation.

## IV. RESULTS AND DISCUSSION

### A. Simulation Results

Typical initial atomic and spin configurations and its evolution after 500 ps of simulation can be seen in Fig. 4 and Fig. 5 for different NP sizes and temperatures. For more insight about the individual atomic spins dynamics, the reader is referred to the movies SM1 and SM2 in the Supplementary Material, where evolution of simulations are shown for a group of spins belonging to a slab at the center of the sphere (SM1) and also for the spins from the surface of the NP (SM2).

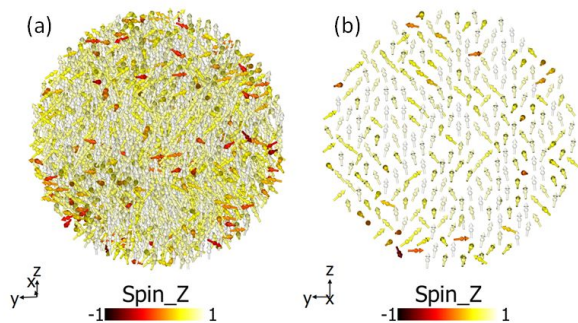

Figure 4. Snapshots of a typical simulation of a NP with a diameter of 6nm. In (a) the atomic spins are displayed as arrows, with color indicating magnitude of $m_z$ at 600 K. (b) Same as (a) for thin slab at the center of the nanoparticle. The initial condition of the simulations was $m_z = 1$, i.e. all spins pointing along the positive z axis. See also the movies SM1 and SM2 in the Supplementary Matterial

The temperature dependence of the total magnetization is shown in Fig. 6 for NPs of 2, 4 and 8 nm in diameter, along with the bulk system. Results for all simulated clusters are not included in this graph to avoid cluttering; a graph with the complete set of the simulated NPs appears in Fig. S2 of the Supplementary Material. Fig. 6 indicates that the bulk Curie temperature, $T_C$, in our simulations is around 650 K, lower than the 1040 K in experiments, and this difference is discussed later on. From Fig. 6 there are indications of an increase of $T_C$ with decreasing cluster size, reaching around $T_C \sim 800$ K for a 2 nm cluster with $\sim 500$ atoms.

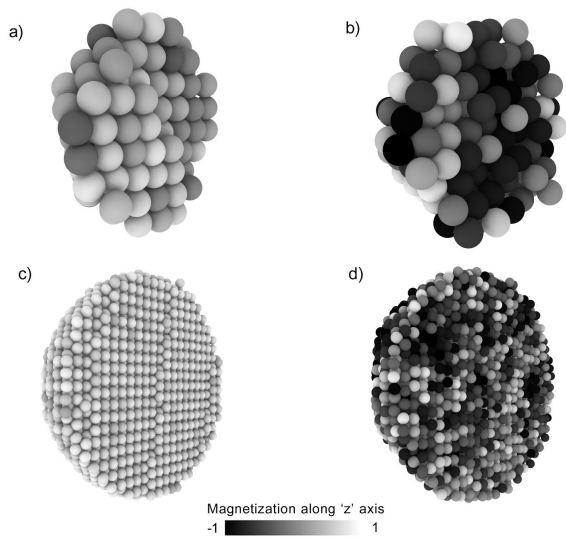

Figure 5. Snapshots of simulations showing the z-component $m_z$ of the atomic magnetization for nanoparticles having (a) 2nm and 300 K, (b) 2nm and 1200 K, (c) 6nm and 300 K and (d) 6nm and 1200 K. The small spheres represent atoms. White indicates that the spins point along the positive z axis, while black indicates that $m_z$ points along the negative z axis. The z axis is vertical with the positive direction pointing upwards in the figures.

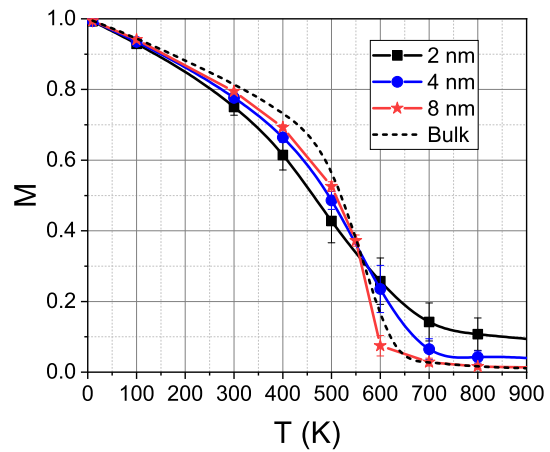

Figure 6. Total normalized magnetization as a function of temperature for different NPs compared to the bulk values.

At the low temperature regime, as the size of the Fe nanoclusters is reduced, the magnetization takes lower values than the bulk ones for all sizes. The difference is larger as the temperature is increased up to about 550-600 K. This is the expected behaviour since the surface contribution to the total magnetization is more important in smaller NPs due to the larger surface/bulk proportion (see Table I). When the temperature is raised up to about T ≈ 500 K, the surface magnetization decreases faster than the core contribution since the spins on the outer layer, having smaller local coordination number, are disordered more easily (see Fig. 11). For temperatures higher than 500 K, the bulk magnetization is the one that decreases faster, i.e changing to the opposite behaviour. This is displayed by the interception of the curves around T ≈ 550 K in Fig. 6. A similar crossover has been also observed in other studies of magnetic nanoparticles [48, 53].

It is interesting to compare our results with those of simulations in which the atoms are static. In Fig. 7 we have compared the magnetization curves with the ones corresponding to nanoparticles with fixed atomic positions. At low temperatures, below 400 K, there is no significant difference. Our results for bulk Fe are consistent with other simulations which found only a small decrease of the critical temperature with the inclusion of a thermalized, moving lattice [36, 54]. For the 2 nm case, the moving atoms lead to a minor decrease in magnetization, since the nanoparticle is already quite disordered, from a magnetic point of view, at the temperatures where the fluctuations of the interatomic distances become important due to the large fraction of surface atoms (see Table I). For the 8 nm case, magnetization goes to zero at lower temperatures for the moving atoms, as expected from the bulk simulations. At 600 K, near the critical temperature, the mean value of the nearest-neighbor (NN) distances changes less than 1% with respect to the one for the frozen lattice. As a result, the value of $J(r_{ij})$ averaged over the distribution of NN atomic distances at 600 K is only slightly different than the value $J(r_{NN})$ for fixed atoms, changing ∼5%. Therefore, neither a lattice expansion nor a fluctuating mean value for $J$ can explain the decrease in magnetization found in our simulations for a moving lattice. This is a clear indication that the coupling between lattice and spins play a significant role for not too small Fe particles near $T_C$. Calculation of time and spatial correlation functions might help understanding this in detail.

A complementary insight is obtained by comparing our results with simulations for static spins in a regular lattice [55]. Let us assume only nearest-neighbor interactions and $J = 15$ meV, similar to the value in our simulations for distances between 1st and 2nd neighbors. The Ising model in 3D gives $T_C = zJ/k_B$ for the mean-field approximation, where z is the coordination number (8 for bcc) resulting in $T_C = 1393$ K. However, Bethe's solution gives $T_C \approx 2J/k_B \ln(z/(z-2)) \approx 1200$ K. The classical Heisenberg model gives $T_C \approx 4zJ/3k_B = 1857$ K, using the mean-field approximation. Using a high temperature expansion, this changes to $(105/96)(z-1)J/k_B = 1333$ K [55]. As expected, the mean-field approximation overestimates the Curie temperature compared to the exact analytical models or to numerical solutions.

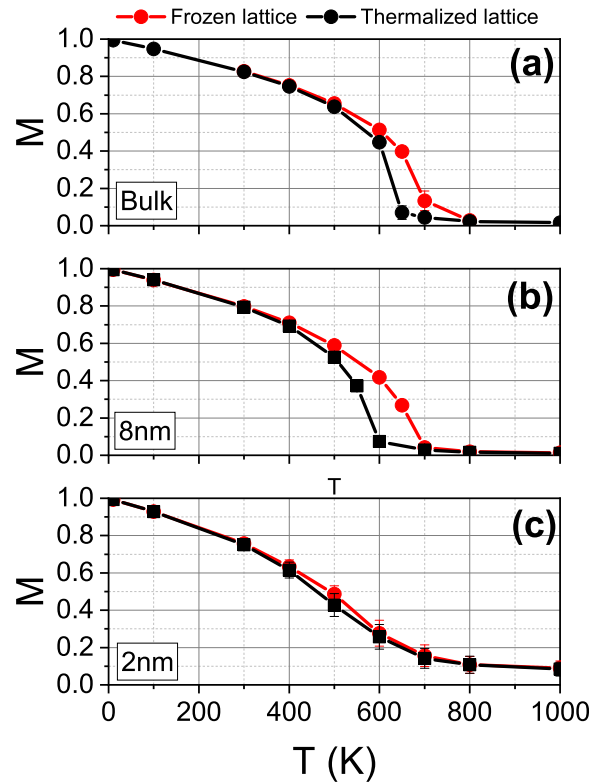

Figure 7. Temperature dependence of the magnetization $M(T)$ as obtained using MD-SD, for (a) Fe bulk, nanoparticle diameters (b) 8 nm, and (c) 2 nm. The results for thermalized and frozen translational degrees of freedom are compared. Bulk results correspond to cubic simulation boxes with $20\, a_0$ sides.

Our NPs magnetization results are consistent, and show really good agreement, with the experimental magnetization curves for Fe nanoclusters reported by Billas *et al.* [56], Fig.8(a), where it can be seen that both, the shape of the curves and the estimated Curie temperature for the iron nanoclusters are well reproduced. To compare with experimental results we have assumed a constant local magnetic moment of $2.2\mu_B$.

While our simulation results match with experiments for small clusters, our bulk simulations yield an estimated Curie temperature around $T_C \approx 650$ K, far below the experimental value of 1043 K. The main reason for this discrepancy is the exchange coupling function used in this work. When $J(r_{ij})$ is replaced, for example, by the one used by Ma *et. al* [30] (note the large discrepancies found in the literature for $J(r)$ reported in Fig.1 of Ref. [30]) our simulations show similar results regarding the experimental Fe Curie temperature, as the results reported in that paper [Fig. 8(b)]. Should one use those parameters for small clusters, one obtains an overestimation of the cluster $T_C$. Hypothetically, this is an indication of an im-portant increase of the effective exchange function $J(r_{ij})$ as the size of the NPs increase. It also reflects the challenge of modelling broad size ranges in the size-dependent effective interaction parameters.

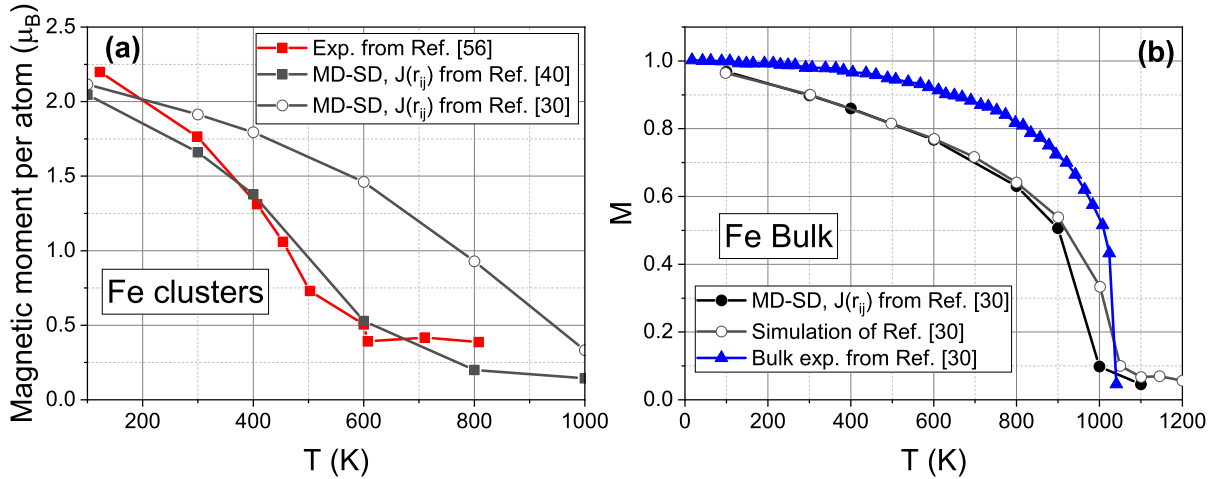

Figure 8. (a) Experimental results for a Fe nanocluster of about 500-600 atoms as reported by Billas *et al.* [56] compared to the results of our simulations of a Fe NP composed of 533 atoms using two different exchange functions $J(r_{ij})$. The exchange function $J(r_{ij})$ fitted from the calculations of Pajda *et al.* [40] is the one used in this work and it reproduce quite closely the experimental results. The open circles correspond to simulation results using the exchange function proposed by Ma *et al.* [30]. (b) Magnetization of a Fe Bulk system (cell size $20 \times 20 \times 20 a_0^3$) as obtained using the exchange function proposed by Ma *et al.* [30] (full circles). Bulk experimental data (triangles) and the results reported by Ma *et al.* (open circles) are also shown for the sake of comparison.

Some recent spin dynamics simulations which incorporate finite temperature effects obtain a lower Curie temperature, $T_C$, than experimental values [36, 57]. It is known that SD simulations tend to smooth the ferromagnetic-paramagnetic transition near the Curie temperature due to the intrinsic coarse-graining of the numerical scheme and since quantum effects are not taken into account [20]. Nevertheless, adding a quantum mechanical treatment does not guarantee a correct determination of $T_C$ [58]. The discrepancy between $T_C$ obtained in our simulations and the experimental one could also be partly related to the fact that our magnetic exchange function $J(r_{ij})$ is not temperature nor size sensitive. In fact, in this framework, it is not possible to take into account any temperature-dependence changes in the electronic structure from which $J(r_{ij})$ is obtained. Modeling an exchange parameter that is function of T is expected to lead to a more precise estimation of $T_C$ [20, 59, 60]. In addition, including system-size variations of $J(r_{ij})$ might help reproducing the changes in $T_C$ which can be derived from Fe cluster experiments [56].

Fig. 9 displays the size dependence of the total magnetization $M(T)$ which is plotted as a function of the inverse diameter for representative temperatures T. This allows to show the bulk magnetization values corresponding to $1/d = 0$. At low temperature, the calculated size dependence is almost negligible. This trend contrasts with the well-known enhancement of the average ground-state magnetization in Fe clusters and surfaces, which can be qualitatively explained from an electronic perspective as a consequence of 3d-band narrowing [61]. The effect could be easily incorporated in our simulation by taking into account the size dependence of the local magnetic moment $\mu_i = |\langle s_i \rangle|$. As the temperature increases, so does the influence of the NP size on the total magnetization. These results are in good agreement with those reported by Iglesias *et. al* [7], who studied the size effects on maghemite nanoparticles via Monte Carlo simulations.

It is notable that Fig. 9 clearly shows two different regimes: low ($T < 600K$) and high ($T > 600K$) temperatures, separated by the 600 K curve. For temperatures below 600 K we find that the magnetization behaviour is almost linear with the inverse diameter, showing higher values for larger nanoparticles. This trend is expected, as adding more magnetic atoms to the particle stabilizes a stronger ferromagnetism, and thus increases the resilience of the total magnetization to thermal disorder. In addition, this linear size-dependence is stronger when the temperature is increased. In order to further analyze these results, we have proposed a functional with only 2 free parameters for the magnetization as a function of temperature and cluster size. This approximation manages to reproduce the size dependence of the magnetization very well for temperatures below 600 K, as it can be seen from the dashed lines in Fig. 9. The corresponding

function is

$$M(T,d) = a(1-b^T)\frac{1}{d} + \left(\frac{T_C - T}{T_C}\right)^{1/3},  \quad (10)$$

with $a = 0.1$, $b = 1.003$, $T_C = 650$ K and $d$ the NPs diameter.

For temperatures above 600 K, the linear relation between $M(T)$ and $1/d$ is reversed: although the trend remains almost linear. At those higger T, larger NPs retain less magnetization than smaller ones. This qualitative change reflects the intersection of the magnetization curves on Fig. 6.

The two different regimes observed in Fig. 9 are separated from each other by the 600 K curve, where a different trend is observed. This different behavior may be due to competing effects, as the system is close to the critical temperature.

Behaviour of the magnetization with varying size is governed by finite-size scaling laws. According to finite-size scaling theory [63, 64], magnetization near the critical point should scale with size as $M \approx L^{(\beta/\nu)}$ where $\beta$ and $\nu$ are the critical exponents related to the order parameter and correlation length respectively. This means that the value of the magnetization for the bulk, near the critical point would be lower than the simple extrapolation from a linear fit, as the one shown in Figure 1. Magnetization near the critical temperature is expected to scale as $M\ (1/L)^{(\beta/\nu)} = (1/L)^{0.514}$, where $\beta/\nu = 0.514$ is the estimated value for the critical exponents of the 3D Heisenberg model [65–68]. Therefore, the bulk value should be closer to the values for finite size nanoparticles, as also shown in that figure. In addition, the behaviour of the 600 K curve for clusters is also consistent with these arguments, as shown by the dashed-dotted line in Fig. 9, corresponding to $M = 0.35(1/d)^{0.514}$.

In practice, Fig. 9 shows that even above $T_c$, small subclusters of atoms with similar spin orientations nucleate in the NPs as well as in the bulk cells (before rapidly dissipating). In large NPs and bulk cells, those local clusters are likely to be averaged down by the larger number of atoms, or by other local clusters with opposition magnetization. As NPs are getting smaller, fewer clusters can nucleate and thus cannot average down the resulting local magnetic moments of each others. This results in the persistence of a small net magnetization at higher temperature. The trend observed above 600 K may be thus explained by the presence of spin-spin correlations above the Curie temperature. Indeed, short-range magnetic ordering (SRMO) is known to persist in the paramagnetic state of iron bulk and surfaces [69–71]. The importance of SRMO in Fe above the Curie temperature has been explicitly demonstrated in Ref. [62], where it was shown that correlated clusters with size $N_{cl}$ for a system with $N$ atoms led to a magnetization $M \propto (N_{cl}/N)^{1/2}$, with $N_{cl} = 15$ for Fe. This dependence is similar to the one above from finite size-scaling, and $N_{cl} = 15$ is consistent with 1st and 2nd NN shells in the bcc lattice, totalling 14 atoms, which is expected since the 2nd NN shell is only around 14% further away than the NN shell, as it is shown later in this section. Fig. 10 shows that our MD simulations near $T_C$, at 700 K, follow this relationship extremely well, and that for larger temperatures $N_{cl}$ decreases significantly, due to thermal fluctuations decreasing magnetic correlation as expected. In future work, the temperature dependence of spin-spin correlation functions and related structure factors could be evaluated for different NP sizes in order to study this in detail.

The two different regimes observed in Fig. 9 may also be explained by assuming distinct core and surface magnetizations. Making this distinction, the total magnetization can be written as $M(T) = M_c(T) - [M_c(T) - M_s(T)]\frac{1}{d}$, where $M_c(T)$ and $M_s(T)$ represents, respectively, the core and shell magnetization contributions for a NP of diameter $d$. In this way, the total magnetization is closer to the bulk magnetization as the size of the NP increases. Consequently, if $M_s(T)$ decreases faster than $M_c(T)$ as the temperature is increased, the slope of $M$ vs. $(1/d)$ becomes steeper, $([M_c(T) - M_s(T)]$ becomes larger), which is the case observed in Fig. 9 until $T = 500\ K$. At 600 K, for the smaller NPs ($\frac{1}{d} \geq 0.33$), both the interior and surface spins are disordered, that is, $M_c(T) \simeq M_s(T)$ and the curve is flatter, i.e. practically does not depend on $1/d$. For larger nanoclusters, both shell and core spins are disordered, resulting in a small global magnetization.

Having a free surface on the NPs naturally introduces finite-size effects on the thermal behaviour of the magnetization. The effects of the free surface were studied for different NPs and results are shown in Fig. 11. Surface (red dots) and core (blue triangles) contributions to the total magnetization (black squares) are plotted along with bulk magnetization results (dashed line). Surface magnetization is calculated averaging over the atoms belonging to a single-atom wide spherical shell composed of the outer layer of atoms as described in sub-section (II-B) (see Fig. 3).

The width of the surface is held constant for all spheres. The main qualitatively result is that the NPs retain less magnetization than the bulk system and this difference in magnetization increases as the particle size decreases. This is a consequence of the low coordination of the spins at the surface, together with the large surface-to-volume ratio and their contributions to the total magnetization. In this way, it is clear that for the 2nm sphere the surface contribution is as relevant as the core one. Therefore, all three curves, $M$, $M_s$ and $M_c$, are far apart from the bulk values. As the particle diameter increases, the surface contribution to the total magnetization decreases. Thus, two main features can be seen: the total magnetization becomes increasingly similar to the core magnetization and, straightforwardly, the core magnetization approaches more and more the bulk magnetization values. This tendency of the core magnetization to approach to the bulk values is stronger at low temperatures, while near $T_C$ there is still a clear departure from the bulk be-haviour. In addition, there is a temperature range where the cluster magnetization behaviour and, in particular, the surface contribution displays an approximately linear temperature dependence, as it has been previously reported by Iglesias *et. al* [7]. Furthermore, this temperature range is larger for smaller particles. In [7], it was argued that this linear behaviour was related to an effective 3D-2D dimensional reduction of the surface shell, and that it has been observed in thin film systems and in simulations of rough ferromagnetic surfaces [72–76].

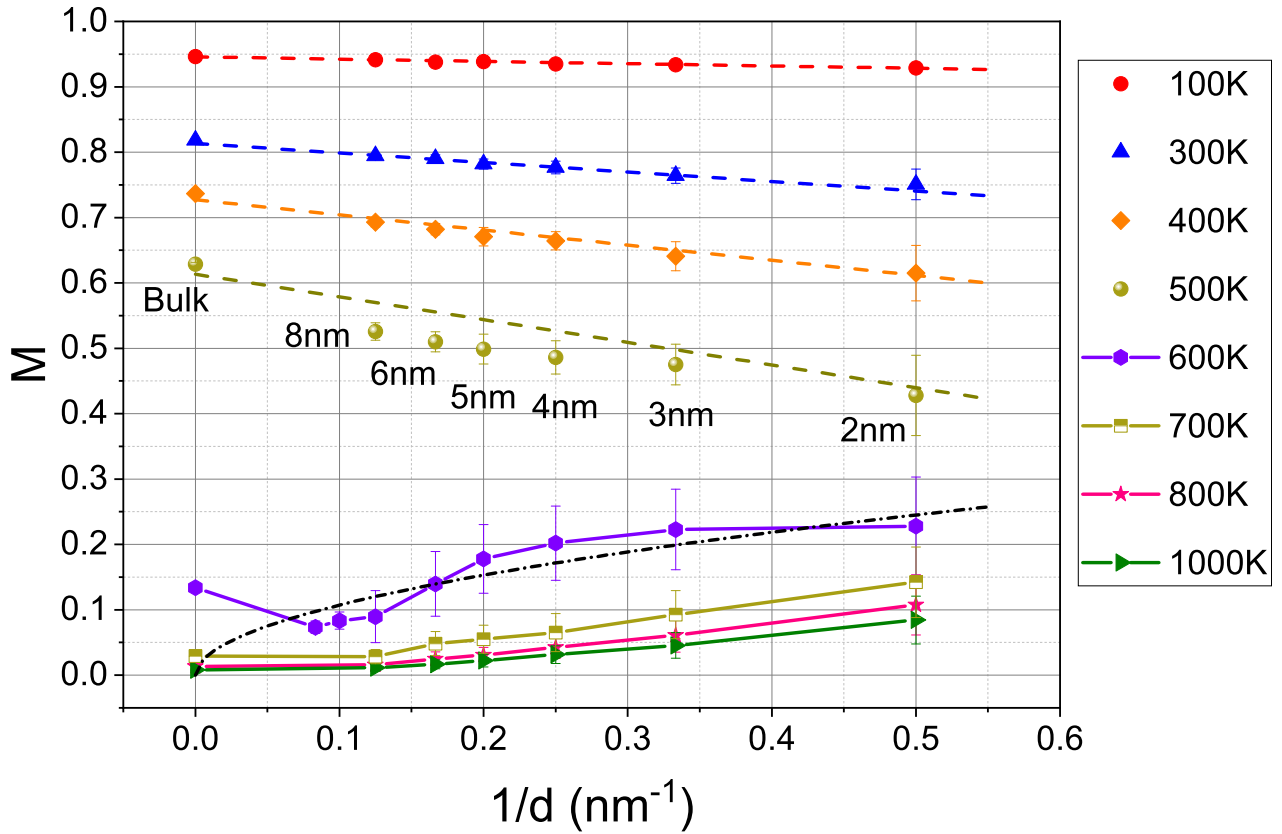

Figure 9. Size dependence of the total magnetization at different representative temperatures. The normalized magnetization is plotted as a function of the particle inverse diameter ($1/d$). Symbols and full curves correspond to the simulation results, whereas the dashed lines represent the fitted curves. The error bars indicate the standard deviation of the data that are taken into account in the average over the last 0.3 ns. For the fitted curves, the function is $M = a(1 - b^T)(1/d) + (\frac{T-T_C}{T})^{1/3}$ where $a = 0.1$, $b = 1.003$, and $T_C = 650$ K. The black dash-dotted line correspond to $M = 0.35(1/d)^{0.514}$, and and is related to the magnetization scaling behaviour near the critical temperature in the 3D Heisenberg model as detailed in the text.

The surface, core and total magnetization are related by

$$M = pM_s + (1-p)M_c, \qquad (11)$$

where $p = N_s/N_T$ is the shell fraction, $N_S$ ($N_T$) being shell (total) number of atoms [7]. Assuming that the shell is much thinner than the nanoparticle radius, one can arrive to the approximate expression

$$M(d) = M_c - \Delta M \frac{\Delta r S}{V} = M_c - \Delta M \frac{6\Delta r}{D} \qquad (12)$$

where $S$ ($V$) are the surface (volume) of the particle, $\Delta r$ is the thickness of the surface layer, $D$ is the diameter of the spheres and $\Delta M = M_c - M_s$. Fig. 12 displays a good agreement between the simple two component approximation and our simulation results. In particular, Eq. (12) manages to reproduce the intersection of the curves around $T_C$, a fact that is related to the surface/core proportion as discussed above.

### B. Exploring possible NP pre-melting

The EAM potential used in the previously mentioned study yields a bulk melting temperature $T_M \simeq 2000K$ [77]. However, small clusters are expected to have much lower $T_M$ due to the reduced surface coordination number. Indeed, in ref. [78], a reduction of about 30% in the melting temperature of 2 nm Fe clusters has been reported. Assuming the same reduction of $T_M$ for the EAM potential used here would give $T_M \simeq 1350K$. The struc-ture and diffusivity of core and shell atoms have been analyzed at different temperatures, to evaluate possible premelting of Fe nanoclusters in the framework of our model. The pair correlation function, $g(r)$, for the whole nanocluster, is shown in Fig. 13. There are broad but well defined peaks for 2nd and 3rd nearest-neighbors at high temperatures, as expected for a crystalline solid.

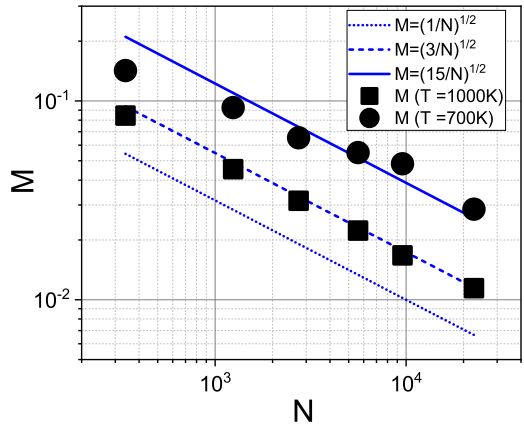

Figure 10. Fe NP magnetization as a function of the number of atoms $N$ in the nanoclusters (symbols) at two representative temperatures above the predicted cluster $T_C$, compared with different power law relations, including the one proposed in Ref. [62], for clusters with 15 atoms.

At 1200 K the nearest-neighbour distance for bulk Fe is 0.245 nm, while for the smallest NPs this distance is closer to 0.25 nm, which correpond to a change of only 2%. The effect of large changes in neighbor distance for Fe films was evaluated by Garibay-Alonso et al [43], who also observed a nearly linear decrease of the layer magnetization with temperature, as in Fig. 11. The radial distribution functions $g(r)$ shown in Fig. 14 for $T = 600$ K, show that there are non-negligible fluctuations on the first and second-neighbor separation distance as the size of the system is reduced. These fluctuations may result in values of $J(r_{ij})$ (see parametrization of $J(r_{ij})$ in [37]), that could drive the systems into an antiferromagnetic phase for the smallest clusters at 600 K, as it is shown in Ghosh et al [79].

In Fig. 14, results are given for the interatomic correlation function $g(r)$ in the shell and core regions at a temperature near the cluster $T_C$. The well-defined peaks in $g(r)$ show no evidence of surface premelting at this temperature, even for the smallest NPs. To further investigate this mater, we have carried out atomic diffusivity calculations. At 600 and 1200 K, diffusivity for core atoms has a value close to zero, but for shell atoms has values of the order of $D = 4.0 \times 10^{-10} m^2/s$, as seen in Figure 15. For reference, the bulk self-diffusion coefficients values are around $D = 1.0 \times 10^{-15} m^2/s$ for 1000 K as reported in [80]. The diffusivity for molten Fe has been typically calculated above 1500 K. For instance, Gosh et al. [81] used the EAM parameters of Bhuiyan et al. [82] and obtained values which, extrapolated down to 1230 K give $D$ close to $1.5 \times 10^{-9} m^2/s$. Given that, according to the results shown in Fig. 15, the shell region of a 4nm Fe NP at 1200 K has a diffusivity similar to extrapolated molten bulk values, one could argue that the shell region is molten or partially molten at 1200 K, while at 600 K it is still solid-like. However, we note that the shell region contains mostly surface atoms, and that surface diffusion is complex at high temperatures [44]. Therefore, further studies are needed to elucidate this point.

### C. Comparison with theoretical models

We have compared our simulation results with a semi-infinite mean-field Ising model, an analytical model that includes surface effects and is described in section III-A. In Fig.16, we have compared the bulk and shell magnetizations obtained from the model with the ones from MD-SD simulations for different values of the exchange constant $J$. As it can be seen, the model results are highly sensitive to the value $J$, but the qualitative behaviour of the system is well reproduced for the three values considered. The thermal behaviour of the analytical surface magnetization reproduces the shape of the corresponding MD-SD curve, also displaying an inflexion point at high temperatures,i.e. $dM/dT$ does not show a monotonous decrease. The rapid decrease of surface magnetization is also observed for the surface magnetization of thin films [43].

Fig.17 shows the analytical results of the model presented in section III-B and the computational results for different NPs, showing a good qualitative agreement. It can be seen that the behaviour of the different magnetizations is well reproduced.

Quantitative agreement is difficult to achieve using mean-field models like the ones in Figs. 16 and 17. As expected, any mean-field model requires an effective J much lower than the one used in our spin Hamiltonian to adjust the critical temperature. Both models show good agreement with our results for $J = 6.5$ meV. In addition, there are non-negligible differences between Ising and Heisenberg models, as expanded below. Nevertheless, these results are usefull in order to test and support the MD-SD results. It is notable how the model manages to reproduce very well the behaviour of the different magnetizations as the size of the nanoparticle is enlarged. It shows, as well as the MD-SD simulations, that an 8nm diameter nanosphere behaves closely to a bulk system.

The Ising model variations shown above do not offer an accurate quantitative prediction of our magnetization curves. All of them use the mean-field approximation, and only up/down spin states. This is because solving the Heisenberg model in 3D is not trivial, even for periodic boundary conditions. Recent work shows the "phase diagram" for different values of $J$ at first, sec-ond and third nearest neighbors ($J_1$, $J_2$, $J_3$) [79]. The frontier for the $(\pi,\pi,\pi)$ antiferromagnetic phase appears at $J_2/J_1 = 2/3 \sim 0.67$. In our case, for the chosen $J(r_{ij})$, this ratio changes with temperature, and it is also affected by strain in the nanoparticle, but is close to $J_2/J_1 = 0.6$, and $J_3/J_1 = 0$. This puts our system in the ferromagnetic state and, therefore, close to the frontier between the $(0,0,0)$ ferromagnetic and $(\pi,\pi,\pi)$ phases. The (q,q,q) spiral phase is close, but $J_3 > 0$ would be needed to reach that region of the phase diagram. At 10 K, the separation distance between nearest-neighbor spins ($d_{nn}$) is around $d_{nn} = 0.245$ nm, and the distance between second nearest-neighbor ($d_{2nd}$) is about $d_{2nd} = 0.285$ nm, resulting in $J_1 = 19.5$ meV and $J_2 = 12.12$ meV, $J_3 = 0$, giving $J_2/J_1 = 0.62$. At 1000 K, $d_{nn} = 0.25$ nm, $d_{2nd} = 0.29$ nm, $J_1 = 19$ meV, $J_2 = 11$ meV, $J_3 = 0$, giving $J_2/J_1 = 0.58$.

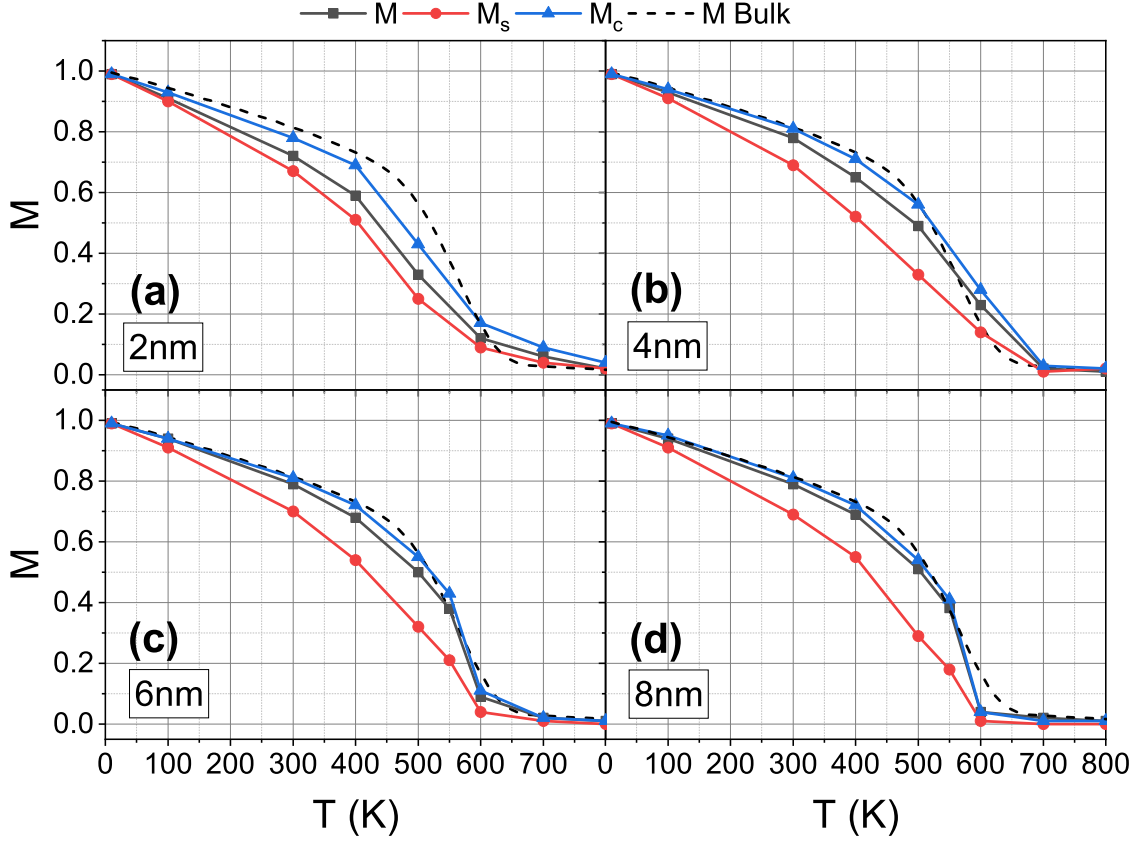

Figure 11. Total normalized magnetization $M$, core magnetization $M_c$ and shell magnetization $M_s$ a functions of temperature for Fe NPs having different diameters. The Fe bulk magnetization is also shown.

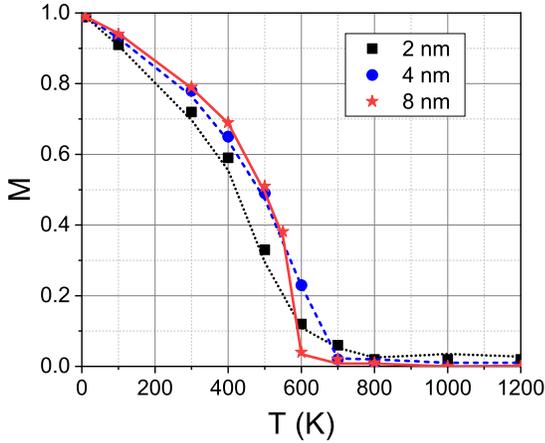

Figure 12. Temperature dependence of the total magnetization $M$ of Fe NPs. The results of our MD-SD simulations (symbols) are compared with the core-shell two component model described by Eq. (12) (curves)

As an alternative simple model for magnetization vs temperature, Fig. S3 of the Supplementary Material shows the curve for the "shape" model by Kuz'min [83]. The comparison of the model with our bulk results shows reasonable agreement if one sets the critical temperature to $T_c = 650$ K in the model.

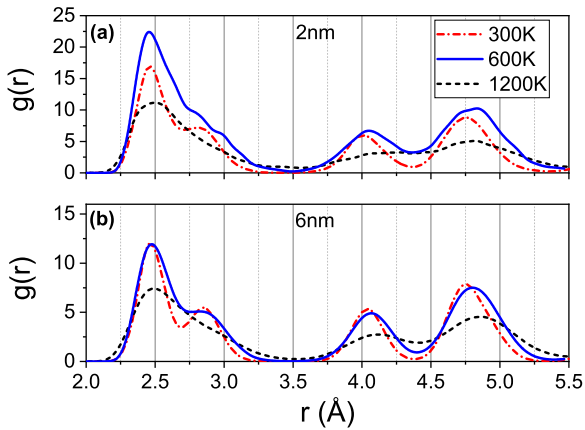

Figure 13. Coordination analysis for spherical Fe NPs with diameters (a) $d = 2nm$ and (b) $d = 6nm$ at different temperatures. The data are taken from the last configuration of the system. They indicate lack of melting, even for the smallest considered nanocluster.

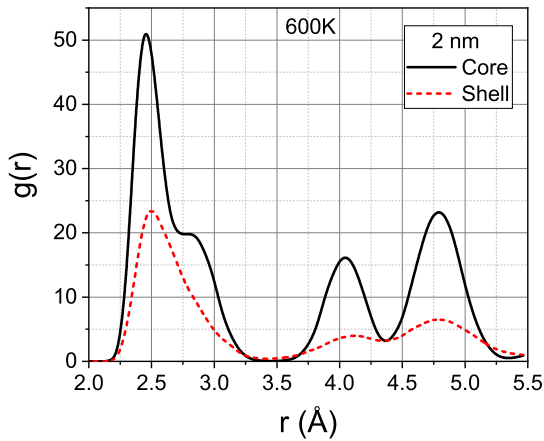

Figure 14. Correlation function in the core and shell regions of a 2nm Fe NP at $T = 1200K$.

## V. CONCLUSIONS

We have performed combined molecular dynamics and spin dynamics simulations on isolated bcc Fe spherical nanoclusters and studied their magnetic properties as a function of temperature and cluster size, in the absence of external magnetic fields. To this aim, we have used a classical spin Hamiltonian coupled to classical molecular dynamics. The effect of anisotropy is also considered. Our results naturally include lattice expansion, surface stress, and other factors, which are difficult to include in spin lattice models. Our simulations include Fe nanoclusters with up to 23000 atoms, and bulk simulations with up to 250,000 atoms in the simulation cell. The temperature of the lattice and the spins can be defined without any additional scaling factors [37], as usually required in

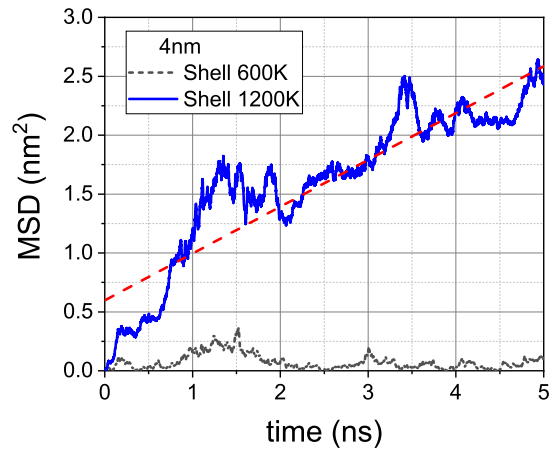

Figure 15. Mean-square displacement of the shell atoms at 600 and 1200 K, as obtained from our simulations for a 4 nm NP. The straight dashed line represents a linear fit of the 1200 K results, giving a diffusion coefficient $D = 4.0 \times 10^{-10} \mathrm{m}^2/\mathrm{s}$. Note that this simulation is 10 times longer (5 ns) than those performed for the magnetization calculations.

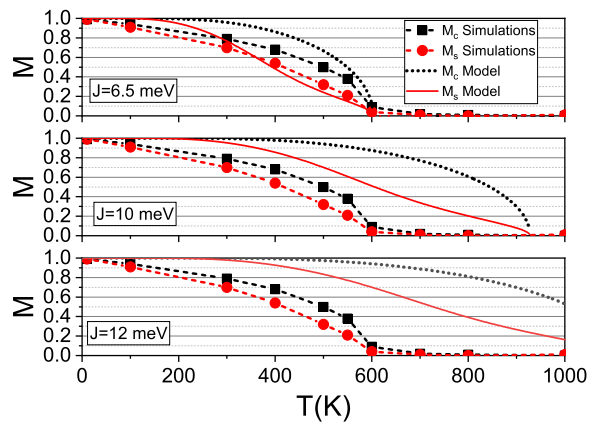

Figure 16. Temperature dependence of the core and shell magnetizations as obtained from the semi-infinite Ising model with a free surface for different values of the exchange constant J (section III.A) compared with MD-SD simulations results. The curves are obtained for a Fe NP with a diameter of 6 nm.

SLD simulations [24].

Given the complexity of solving 3D magnetic models, simulations including thermal lattice effects, like thermal expansion and lattice strain due to surface effects, can be valuable tools in understanding and predicting the behavior of nanoscale magnetic systems. We find significant differences between our simulations with moving atoms, and simulations with frozen atoms as in most atomistic spin-dynamics simulations, specially at temperatures above $2/3$ of the critical temperature. Recently, lattice relaxation was found to produce significantly larger coercivity enhancement in nanoparticles

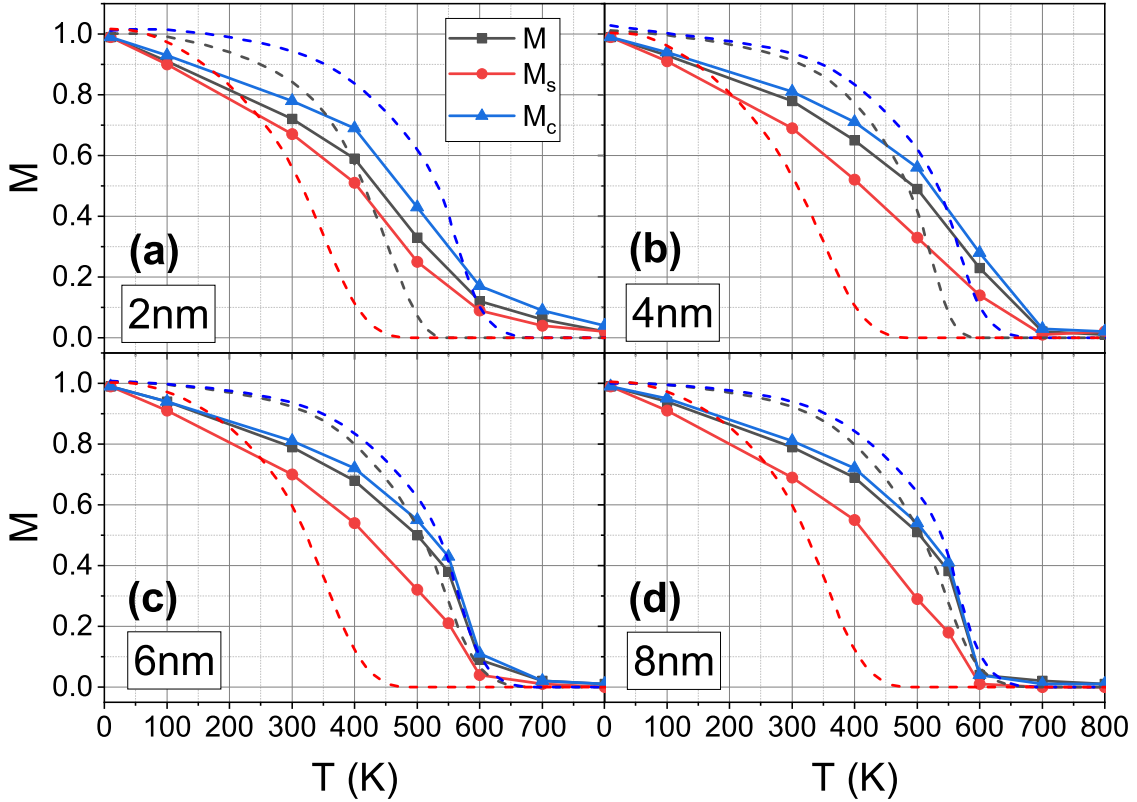

Figure 17. Total, core and shell magnetization curves $M$, $M_c$, $M_s$, as obtained in the model presented in section III.B (dashed lines) [51] and in the present MD-SD simulations (solid line and symbols) for Fe NPs having different diameters.

than the case of an unrelaxed, fixed lattice [84].

Our results show excellent agreement with experimental measurements of Fe nanoclusters [56]. The magnetization thermal behaviour of small clusters is well reproduced and the estimated Curie temperature is also very similar. Total magnetization decreases with increasing temperature and decreasing size and, therefore, the decrease of magnetization with temperature is faster for the smallest clusters. Qualitatively, these results are expected, but they are quantitatively different from the ones in simple semi-analytical mean-field Ising models, even when size and surface effects are considered. Above the Curie temperature we find that magnetization scales with system size as predicted by models assuming short-range magnetic ordering (SRMO) [62].

For temperatures above 1000 K, we observe evidence of possible melting of the shell region, as shown by both diffusivity and radial coordination studies. This is consistent with melting temperature reduction due to finite size and surface effects in nanoclusters, but only occurs well above $T_C$ in our model.

Regarding our bulk simulations, we obtain an estimated Curie temperature close to $T_C = 650$ K for bulk systems. The discrepancy with the experimental value is attributed to the exchange coupling $J(r_{ij})$. Large differences of the reported values of $J$ as a function of pair separation distance are found in the literature and, therefore, fitting the function $J(r_{ij})$ with a different set of ab-initio calculations results in different magnetization curves. This statement is clearly verified in Fig.8 where the bulk Curie temperature is well reproduced if a different set of ab-initio data is used to fit the $J(r_{ij})$ function. Our calculations indicate that possible size-dependence of $J(r_{ij})$ might lead to significant magnetization changes.

The total NP magnetization can be considered to be the result of an ordered core plus a less ordered outer shell. In fact, simple two-component models provide a reasonable fit to our results. Core magnetization resembles bulk magnetization and, as cluster size increases, dominates the overall magnetization. Shell magnetization is significantly lower than bulk magnetization, as expected due to the reduction of the local coordination number.

We propose a functional form for the magnetization as a function of size and temperature, which has only 2 free parameters and works extremely well for temperatures below $T_C$, and for the range of sizes studied, from 2 nm cluster diameter up to bulk conditions.

It is clear that the classical MD-SD simulations presented in this work still have several limitations, for example, assuming classical continuous spin variables, or the fact that exchange, anisotropy, and magnetic moments are the same for surface and core atoms. However, they are expected to contribute to the understanding of magnetism in nanoscale systems, providing quantitative values to compare with experiments for nanocluster magnetization. Among the improvements to be implemented in future studies, it would be interesting to consider the effect of defects in the clusters, such as vacancies, impurities, dislocations and grain boundaries, together with the role of an external magnetic field and dipolar interactions with other nanoclusters.

## ACKNOWLEDGMENTS

We thank Federico Romá for helpful discussionss. EMB thanks support from the SIIP-UNCuyo grant 06/M104. We thank computer run time in the cluster TOKO (toko.uncu.edu.ar). The authors thank support from IPAC-2019 grant from "Sistema Nacional de Computación de Alto Desempeño" (SNCAD) for run time in the cluster Dirac http://dirac.df.uba.ar/. Sandia National Laboratories is a multimission laboratory managed and operated by National Technology & Engineering Solutions of Sandia, LLC, a wholly owned subsidiary of Honeywell International Inc., for the U.S. Department of Energy's National Nuclear Security Administration under contract DE-NA0003525. This paper describes objective technical results and analysis. Any subjective views or opinions that might be expressed in the paper do not necessarily represent the views of the U.S. Department of Energy or the United States Government.

## SUPPLEMENTARY MATERIAL

### Supplementary Movies

Follow this link : Supplementary Movie 1

SM1. Spin dynamics for a slice at the center of the 4 nm wide sphere held at 600 K. Small spheres represent atoms and arrows indicate spin orientation. Both are color coded according to the z component of the atomic spin.

Follow this link : Supplementary Movie 2

SM2. Spin dynamics for the outer shell of the 4 nm wide sphere held at 600 K showing half of the sphere. Small spheres represent atoms and arrows indicate spin orientation. Both are color coded according to the z component of the atomic spin.

### Supplementary Figures

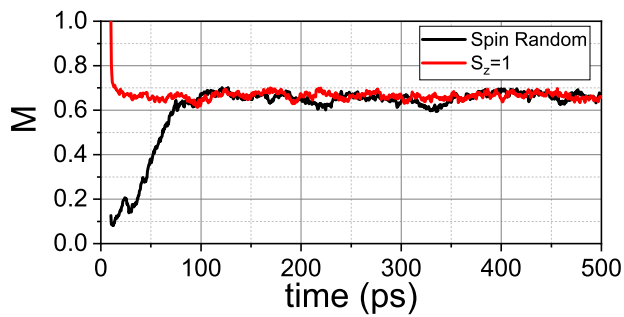

Figure S1. Normalized magnetization at 400 K for a NP with a diameter of 4 nm, as a function of simulation time for different initial simulation setups: All spins initially oriented along the z axis (red), and spins oriented randomly (black).

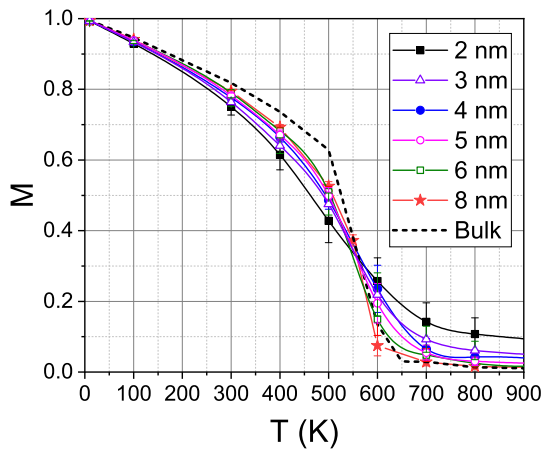

Figure S2. Total normalized magnetization as a function of temperature. Same as Fig. 6 of the main article, but for all the different NPs compared to the bulk values.

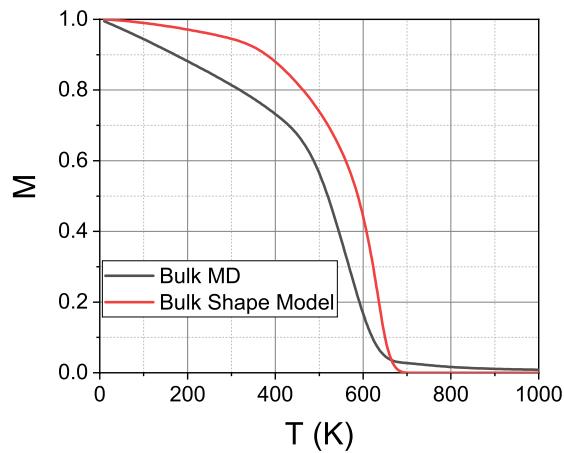

Figure S3. Bulk magnetization obtained from MD-SD simulations compared to the semi-empirical model developed by Kuz'min et al. [83]. The parameters of the model curve are taken from the mentioned paper as well.